\documentclass[10pt,twocolumn]{revtex4}
\usepackage{amsmath}
\usepackage{amssymb}
\usepackage{comment}

\newcommand\beq{\begin{equation}}
\newcommand\eeq{\end{equation}}
\newcommand\bea{\begin{eqnarray}}
\newcommand\eea{\end{eqnarray}}
\newcommand\nn{\nonumber}

\newcommand{\bra}[1]{\langle #1 |}
\newcommand{\ket}[1]{| #1 \rangle }

\newcommand{\ga}{\gamma}

\newcommand{\Si}{\Sigma}
\newcommand{\si}{\sigma}

\newcommand\Hcal{\mathcal{H}}
\newcommand\Hcalo{\hat{\mathcal{H}}}
\newcommand\Ncalo{\hat{\mathcal{N}}}
\newcommand\Ocalo{\hat{\mathcal{O}}}

\newcommand\PSi{\ket{\Psi,\Si}}
\newcommand\PSI{\ket{\Psi_\Si}}
\newcommand\Uo{\hat{U}}
\newcommand\Oo{\hat{O}}

\begin{document}
\title{Comment on ``Relativistic covariance and nonlinear quantum mechanics: Tomonaga-Schwinger analysis'' }
\author{Lajos Di\'osi}
\affiliation{Wigner Research Center for Physics, H-1525 Budapest 114 , P.O.Box 49, Hungary}
\affiliation{E\"otv\"os Lor\'and University, H-1117 Budapest, P\'azm\'any P\'eter stny. 1/A, Hungary}
\date{\today}

\begin{abstract}
Contrary to the central claim (Hsu, 2026) published in Physics Letters B,
the Tomonaga--Schwinger equation remains covariant despite
the nonlinear modification of it. The proof of covariance becomes simple after  
the loopholes and mistakes in Hsu's arguments are identified. 
\end{abstract}

\maketitle
In a recent paper published in Physics Letters B \cite{hsu2026relativistic}, Hsu considered the nonlinear 
Tomonaga-Schwinger (TS)  equation and concluded that the state-dependent Hamiltonian term violates
the relativistic covariance. For standard linear TS equation, the integrability conditions ensure foliation 
independence and relativistic covariance.
The author assumes correctly that also the covariance of nonlinear TS equation is ensured by its
integrability conditions, but argues incorrectly that they are not satisfied because of the breakdown of
microcausality.
We revisit Hsu's analysis, pinpoint its loopholes, contradictions, and its misunderstanding of the 
mechanism of causality violation in nonlinear quantum dynamics.  In our corrected analysis, the TS equation is
integrable and ensures covariance, despite the loss of causality in nonlinear quantum dynamics.  

Let us follow  ref. \cite{hsu2026relativistic}. The nonlinear TS equation reads:
\beq\label{TS}
i\hbar\frac{\delta}{\delta\si(x)}\PSi=(\Hcalo(x)+\Ncalo_x[\Psi])\PSi,
\eeq
with the condition of integrability: 
\beq\label{cond}
\left[\frac{\delta}{\delta\si(x)},\frac{\delta}{\delta\si(y)}\right]\PSi=0,
\eeq   
which is the condition of foliation independence, i.e., of covariance.

The concrete choice of nonlinearity  is the specific operator-expectation nonlinearity:
\beq\label{Weinberg}
\Ncalo_x[\Psi]=\lambda\Ocalo(x)\bra{\Psi,\Si}\Ocalo(x)\PSi.
\eeq
The integrability conditions are satisfied if the fields $\Hcalo,\Ocalo$ satisfy the conditions of
microcausality, i.e., they commute at spacelike separated $x$ and $y$.  In the standard linear TS,
the fields evolve unitarily and commute at spacelike separations. Here, let they evolve with the
nonlinear Hamiltonian. If the evolution of the state from $\Si_0$ to $\Si$ is written in
 the schematic form 
\beq\label{U}
\PSi=\Uo[\Psi;\Si,\Si_0]\PSi_0,
\eeq
then the transformation $\Uo$ will transform the fields. Hsu observes that
$\Uo$ is `not unitary in the usual sense`, and concludes that
microcausality cannot be ensured, nor can the  integrability.

We show, however, that $\Uo$ preserves the commutators of $\Ocalo(x)$.
It is true that the transformation  (\ref{U}) of $\PSi_0$ is not unitary. But
the transformation of the field, i.e.:
 \beq\label{UOU}
\Uo[\Psi;\Si,\Si_0]\Ocalo(x)\Uo[\Psi;\Si,\Si_0]^\dagger,~~~~x\in\Si_0,
\eeq
is unitary because $\Uo$ itself is  a unitary operator as long as we apply it
to a $\PSi_0$-indepedent object, vector or operator like $\Ocalo$ above. 
[Note in passing that eq. (\ref{UOU}) assumes zero shifts between 
internal (spatial) coordinates of foliations, the trivial one is the Minkovski foliation. 
The general form would best be written for the differential evolution  
$\delta\Ocalo_\Si(x)/\delta\si(y)$.]
We might perhaps prove the integrability within Hsu's approach to TS eq. (\ref{TS}),
had it not contained a further inconsistency. It is not clear why the fields must evolve with the same transformation 
$\Uo$, which evolves the state. We take Hsu's example of standard (linear) field
theory in Minkowski foliation and evaluate the expectation value of $\Ocalo$:
\bea
\bra{\Psi,t}\Ocalo(x,t)\ket{\Psi,t}&=&\bra{\Psi,t_0}\Uo^\dagger(t,t_0)\Ocalo(x,t)\Uo(t,t_0)\ket{\Psi,t_0}\nn\\
                                                                   &=&\bra{\Psi,t_0}\Ocalo(x,t_0)\ket{\Psi,t_0}.
\eea
The result is the t-independent constant initial expectation value. 
(We took the standard transformation $\Uo\Ocalo\Uo^\dagger$ instead of the  certainly mistaken  
form $\Uo^\dagger\Ocalo\Uo$ in ref. \cite{hsu2026relativistic}.) In the nonlinear case we obtain
a similar result, and the nonlinear Hamiltonian density becomes 
\beq
\Ncalo_{x,t}[\Psi]=\lambda\Ocalo(x,t)\bra{\Psi,t_0}\Ocalo(x,t_0)\ket{\Psi,t_0},
\eeq
in Minkovski foliation. This is a highly degenerate and strange structure of nonlinearity, definitely not the 
desired form. The suggested combination of field and state evolutions are obviously not correct in ref. \cite{hsu2026relativistic}.   

The above problems would have not arisen if Hsu had interpreted the TS eq. (\ref{TS}) in the standard interaction picture of
field theory. The precise form of the TS equation (of the state vector $\PSi\equiv\PSI$) reads:
\bea
\label{TSE}
i\hbar\frac{\delta}{\delta\si(x)}\PSI&=&\Hcalo_\Si(x,O_\Si(x))\PSI,~~~~~~(x\in\Si)\nn\\
                                                        O_\Si(x)&=&\bra{\Psi_\Si}\Oo(x)\PSI.
\eea
$\Hcalo_\Si$ is the interaction Hamiltonian's spatial density on $\Si$, and the quantized local fields evolve
with the free Hamiltonian. The interaction Hamiltonian density at spacetime location $x$ may depend
on the expectation value of the field $\Ocalo$ at the same location. That is, the nonlinearity of the 
Hamiltonian {\it respects locality}. Without the nonlinearity, the TS equation is integrable in local field theory.
In the presence of the operator-expectation nonlinearity, its $\Si$ dependence would yield an extra term
to the integrability condition but, fortunately, it cancels because its derivative cancels already:  
\bea
&&\frac{\delta}{\delta\si(y)}\bra{\Psi_\Si}\Oo(x)\PSI=\nn\\
&&=-i\hbar\bra{\Psi_\Si}\left[\Hcalo_\Si(y,O_\Si(y)),\Oo(x)\right]\PSI=0.
\eea
Here we applied the TS eq. (\ref{TSE}) and the microcausality condition which guarantees that
$\Hcalo_\Si$ and $\Oo$ commute at spacelike separations. This completes our proof that in local
field theory the nonlinearity of the TS equation does not violate integrability and covariance.

To support the contrary, 
Hsu invokes previous works \cite{ho2015locality,gisin1990} on causality violation in nonlinear quantum dynamics.
In the author's understanding,  nonlinearity causes the breakdown of microcausality because
an initially unentangled state $\ket{\psi_A}\otimes\ket{\psi_B}$ of spacelike separated
subsystems $A$ and $B$ becomes instantaneously entangled. While really true in ref. \cite{ho2015locality},
it is absolutely wrong in our case. In ref. \cite{ho2015locality} the chosen nonlinearity was already nonlocal.  
In our case, i.e., in Hsu's own specific example (\ref{Weinberg}) as well as  in our general form (\ref{TSE}),
the nonlinearity was local. We emphasised that nonlinearity (\ref{TSE}) respected locality, a natural desire in local field theories.
Hsu's further argument is based on the essential misunderstanding of Gisin's seminal paper 
\cite{gisin1990}. It proved that if the spacelike separated systems $A$ and $B$ are entangled initially
then any \emph{local nonlinearity} allows superluminal communication via standard local measurements
on $A$ and $B$. Local linearities do not generate entanglement. Causality violation assumes initial entanglement,
local nonlinearity, and not least  the insistence on quantum state collapse mechanism.
The mechanism of acausality is completely different from what ref. \cite{hsu2026relativistic} says about it.

Let us summarize our comments. First, the central claim of Hsu in ref. \cite{hsu2026relativistic} is incorrect because the
nonlinear TS equation is integrable and its covariance is thus preserved. We proved it easily by avoiding the
redundant and problematic mix of state and field dynamics, respectively, in favor of the standard interaction picture. 
Second,  ref. \cite{hsu2026relativistic} misinterprets the relationships among quantum nonlinearity, entanglement generation, 
acausality, and covariance.  Quantum nonlinearity, unless it is nonlocal by itself, does not generate entanglement,
it causes acausality though, but it is prompted by local measurements instead of microcausality violation. 
Loss of covariance cannot be derived from loss of causality caused by nonlinear quantum dynamics.
Third, Hsu's claim is tenuous from the perspective of semiclassical gravity \cite{moller1962,rosenfeld1963},
which is our unique theory encompassing a nonlinear TS equation. Its integrability and covariance would be
impossible if ref.  \cite{hsu2026relativistic} were correct.
  
{\bf Acknowledgements:}  
This research was supported  by the
National Research, Development and Innovation Office
''Frontline'' Research Excellence Program (Grant No.
KKP133827), and by the EU COST Actions (Grants CA23115, CA23130). 
I thank Stephen Hsu for a helpful correspondence. 
\bibliography{diosi2024}{}

\begin{thebibliography}{5}
\expandafter\ifx\csname natexlab\endcsname\relax\def\natexlab#1{#1}\fi
\expandafter\ifx\csname bibnamefont\endcsname\relax
  \def\bibnamefont#1{#1}\fi
\expandafter\ifx\csname bibfnamefont\endcsname\relax
  \def\bibfnamefont#1{#1}\fi
\expandafter\ifx\csname citenamefont\endcsname\relax
  \def\citenamefont#1{#1}\fi
\expandafter\ifx\csname url\endcsname\relax
  \def\url#1{\texttt{#1}}\fi
\expandafter\ifx\csname urlprefix\endcsname\relax\def\urlprefix{URL }\fi
\providecommand{\bibinfo}[2]{#2}
\providecommand{\eprint}[2][]{\url{#2}}

\bibitem[{\citenamefont{Hsu}(2026)}]{hsu2026relativistic}
\bibinfo{author}{\bibfnamefont{S.}~\bibnamefont{Hsu}},
  \bibinfo{journal}{Physics Letters B} \textbf{\bibinfo{volume}{872}},
  \bibinfo{pages}{140053} (\bibinfo{year}{2026}).

\bibitem[{\citenamefont{Ho and Hsu}(2015)}]{ho2015locality}
\bibinfo{author}{\bibfnamefont{C.~M.} \bibnamefont{Ho}} \bibnamefont{and}
  \bibinfo{author}{\bibfnamefont{S.~D.} \bibnamefont{Hsu}},
  \bibinfo{journal}{International Journal of Modern Physics A}
  \textbf{\bibinfo{volume}{30}}, \bibinfo{pages}{1550029}
  (\bibinfo{year}{2015}).

\bibitem[{\citenamefont{Gisin}(1990)}]{gisin1990}
\bibinfo{author}{\bibfnamefont{N.}~\bibnamefont{Gisin}},
  \bibinfo{journal}{Physics Letters A} \textbf{\bibinfo{volume}{143}},
  \bibinfo{pages}{1} (\bibinfo{year}{1990}), ISSN \bibinfo{issn}{0375-9601},
  \urlprefix\url{http://www.sciencedirect.com/science/article/pii/037596019090786N}.

\bibitem[{\citenamefont{M{\o}ller}(1962)}]{moller1962}
\bibinfo{author}{\bibfnamefont{C.}~\bibnamefont{M{\o}ller}},
  \bibinfo{journal}{Colloques Internationaux CNRS}
  \textbf{\bibinfo{volume}{91}}, \bibinfo{pages}{353} (\bibinfo{year}{1962}).

\bibitem[{\citenamefont{Rosenfeld}(1963)}]{rosenfeld1963}
\bibinfo{author}{\bibfnamefont{L.}~\bibnamefont{Rosenfeld}},
  \bibinfo{journal}{Nuclear Physics} \textbf{\bibinfo{volume}{40}},
  \bibinfo{pages}{1} (\bibinfo{year}{1963}).

\end{thebibliography}
\end{document}